\documentclass[prl,reprint,superscriptaddress,aps,floatfix]{revtex4-2}
\usepackage{amsmath}
\usepackage{amsfonts}
\usepackage{graphicx}
\usepackage{physics}
\usepackage[usenames]{color}
\usepackage{braket}
\usepackage{booktabs}

\begin{document}

\date{\today}

 \title{The Floquet central spin model: A platform to realize time crystals, entanglement steering, and multiparameter metrology}
\author{Hillol Biswas}
\author{Sayan Choudhury}
\email{sayanchoudhury@hri.res.in}
\affiliation{Harish-Chandra Research Institute, Chhatnag Road, Jhunsi, Allahabad 211019, India}
\affiliation{Homi Bhabha National Institute, Training School Complex, Anushakti Nagar, Mumbai 400094, India}

\begin{abstract}
We propose and characterize protocols to engineer exact quantum revivals, discrete time crystals (DTCs), and entanglement oscillations in the periodically driven central spin model. While period-doubling DTCs have been observed in this system before, we uncover a unifying interaction-induced echo mechanism underlying several distinct dynamical regimes. We first demonstrate that this echo can enable exact period-doubling revivals when the Ising interaction strength, $\lambda$, between the central spin and the $N_{\rm sat}$ satellite spins is tuned to $2 \pi$. Notably, these revivals persist for arbitrary $N_{\rm sat}$ and transverse field $g$. This not only stabilizes the DTC response over a wide parameter regime, but also leads to a dynamical freezing regime. Furthermore, when $\lambda=(2m+1)\pi$ and $g= (2n+1)\pi/2$ ($\forall \, m,n \in \mathbb{Z}$), this echo induces a Clifford group structure. Consequently, higher-period revivals emerge that naturally steer the system through an entangled manifold of Bell-cat and spin-cat states. We establish that due to parity-dependent phases accumulated during the echo, the recurrence period is $12 T\,\, (24 T)$ for even (odd) $N_{\rm sat}$. Finally, we demonstrate that the multipartite entanglement generated by the Clifford dynamics can be harnessed for multiparameter metrology, with odd $N_{\rm sat}$ enabling Heisenberg-limited sensitivity.
\end{abstract}

\maketitle

\emph{Introduction:} Periodically driven (Floquet) quantum systems host non-equilibrium phases of matter with no equilibrium analog~\cite{bukov2015universal,harper2020topology,rudner2020band,oka2019floquet,weitenberg2021tailoring,banerjee2024emergent}. One of the most striking examples of a Floquet phase of matter is a discrete time crystal (DTC)~\cite{khemani2016phase,yao2017discrete,else2016floquet,von2016absolute,sacha2015modeling}. DTCs are characterized by discrete time-translation-symmetry-breaking (TTSB) and a consequent sub-harmonic response of physical observables~\cite{yao2018time,else2020discrete,zaletel2023colloquium,sacha2017time}. Interestingly, despite their non-equilibrium nature, all known DTCs exhibit no growth of entropy, after an initial transient. This phenomenon has been dubbed as crypto-equilibrium~\cite{yao2018time}, and it indicates that the system is in equilibrium in an appropriate rotating frame~\cite{else2020discrete}.  Apart from its fundamental importance as a non-equilibrium phase of matter, DTCs are also expected to be useful for applications in quantum metrology~\cite{estarellas2020simulating,moon2024discrete,lyu2020eternal,yousefjani2024discrete,iemini2024floquet,shukla2025prethermal}. DTCs have now been experimentally realized in various platforms including trapped ion chains~\cite{zhang2017observation,kyprianidis2021observation}, Carbon-13 nuclear spins in diamond~\cite{choi2017observation,randall2021many,beatrez2023critical}, Rydberg atom arrays~\cite{bluvstein2021controlling,liu2024higher}, superconducting quantum processors~\cite{mi2022time,bao2024creating,xu2021realizing,frey2022realization}.\\
\begin{figure}[ht!]
\centering
    \includegraphics[width=0.45\textwidth]{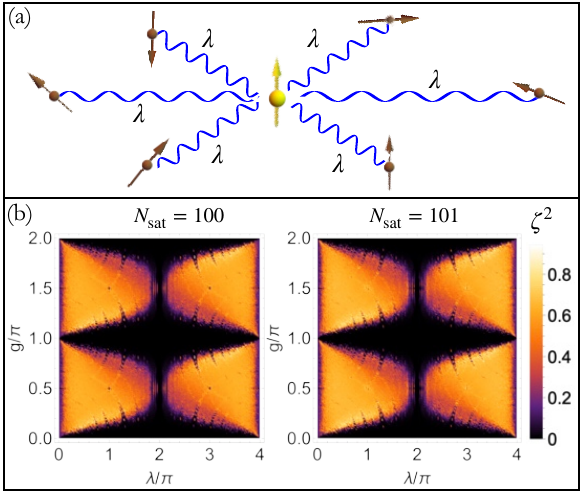}
\caption{{\bf Model and period-blind phase diagram:} (a) A schematic illustration of the central spin model where one central spin-$1/2$ particle is coupled to $N_{\rm sat}-$ non-interacting satellite spin-$1/2$ particles. Each satellite spin interacts with the central spin with strength $\lambda$. (b) Density plot of the period-blind order parameter, $\zeta^2$ (Eq.~\ref{eq:ChaosOP}) for $N = 500$ and $\delta g = \delta  \lambda = 0.001$. DTCs are characterized by small values of $\zeta^2$. A wide parameter regime corresponding to the period-doubling DTC phase is observed.}
    \label{fig:schematic}
\end{figure}
Interactions are crucial for stabilizing TTSB and realizing a robust DTC~\cite{khemani2019brief,sacha2020time}. Unfortunately, interacting Floquet systems generically heat up to a featureless temperature state~\cite{d2014long,lazarides2014equilibrium,choudhury2014stability,zhang2015thermalization}, thereby posing a major challenge to realizing a DTC~\cite{moessner2017equilibration}. Consequently, the initial realization of DTCs relied crucially on many-body localization to evade Floquet heating~\cite{zhang2017observation}. However, in later years, several studies have demonstrated the possibility of realizing a DTC without disorder~\cite{rovny2018observation,pal2018temporal,rovny2018p,huang2018clean,pizzi2019period,pizzi2021higher,munoz2022floquet,russomanno2017floquet,surace2019floquet,maskara2021discrete,sarkar2024time,liu2023discrete}. One of the most celebrated examples of such a disorder-free DTC is in the Floquet central spin model (CSM), where long-lived period-doubling oscillations of the magnetization have been theoretically predicted~\cite{kumar2024hilbert,frantzeskakis2023time,cabot2022metastable} and experimentally observed~\cite{pal2018temporal,geng2021ancilla}. However, the full dynamical phase diagram of this model, and its possible applications for quantum metrology, have not been explored.\\
In this Letter, we address this gap by introducing an interaction-induced echo protocol that can provide a unifying mechanism to realize several non-equilibrium phenomena. In particular, we identify parameter regimes where the echo leads to exact state revivals at times $2T$, $12T$, and $24T$. The robustness of the period-$2T$ revivals leads to the emergence of a period-doubling DTC and a dynamical freezing regime. Interestingly, the higher-period revivals emerge, when the echo induces a Clifford group structure. The parity-dependent phase accumulation in this regime leads to oscillation periods of $12T$ and $24T$ for even and odd values of the satellite spin number, $N_{\rm sat}$, respectively. The protocol naturally generates maximally entangled Bell-cat~\cite{vlastakis2015characterizing} and spin-cat~\cite{dooley2013collapse} states during the dynamical evolution. Finally, we demonstrate that the Clifford-induced multipartite entanglement can be harnessed for quantum-enhanced sensing of the interaction and magnetic field strength, achieving Heisenberg-limited sensitivity when $N_{\rm sat}$ is odd.

\emph{Model and non-ergodic dynamics:} We explore the dynamics of a periodically driven CSM composed of a central spin ($S_c$) interacting with $N_{\rm sat}$ satellite (or ancilla) spins ($S_i$) with an Ising interaction $\lambda$:  
\begin{equation}
H(t) =
     \begin{cases}
			H_d= & -\frac{2}{T}{} (g_s\sum_{i=1}^{N_{\rm sat}} S_i^z+g_c S_c^z) {\rm~for~} t\in[0,~T/2)\\
			H_0=& -\frac{2\lambda}{T} \sum_{i=1}^{N_{\rm sat}} S_i^x S_c^x {\rm~for~} t\in[~T/2,~T).
      \end{cases}
    \label{eq:Hamiltonian}
\end{equation}
Here $S^{\gamma} = \frac{1}{2} \sigma^{\gamma}$, where $\sigma^{\gamma}$ are the Pauli matrices, and $g_c$ and $g_s$ are the strengths of the magnetic fields acting on the central and satellite spins, respectively; we set $g_s=g_c=g$ in our calculations. A schematic illustration of the model is shown in Fig.~\ref{fig:schematic}(a). This model can be realized in quantum dots~\cite{urbaszek2013nuclear}, ultracold atomic gases~\cite{ashida2019quantum,dobrzyniecki2023quantum}, nitrogen-vacancy centers in diamond~\cite{schwartz2018robust,childress2006coherent}, and solid-state nuclear magnetic resonance~\cite{niknam2021experimental}.\\
We begin our analysis of this system by identifying the parameter regimes where the system exhibits non-ergodic dynamics. To this end, we compute a period-blind order parameter that has been employed to detect DTCs~\cite{giachetti2023fractal}:
\begin{equation}
\zeta^2 = \frac{1}{N}\sum_{n=1}^{N} \Big[O_n(\lambda,g) - O_n(\lambda+\delta\lambda,g+\delta g)\Big]^2,
\label{eq:ChaosOP}
\end{equation}
where $O_n (\lambda,g) = \langle \psi(nT) \vert \hat{O} \vert \psi(nT) \rangle$; our results are shown in Fig.~\ref{fig:schematic}(b) for $\hat{O} = \frac{1}{N_{\rm sat}}\sum_{i=1}^{N_{\rm sat}} \sigma_i^x$. We note that $\zeta^2$ is essentially the time-averaged `decorrelator' that has been employed to characterize chaotic dynamics in driven long-range interacting systems~\cite{pizzi2021higher,sinha2024classical}. For chaotic dynamics, $\zeta^2$ grows rapidly due to its salient sensitive dependence on initial conditions, and it saturates to a high value (peaked around $2/3$). For DTCs, on the other hand, the system hops between regular islands in phase space, leading to a small value of $\zeta^2$~\cite{giachetti2023fractal}. Our results clearly indicate a wide parameter regime where $\zeta^2 \approx 0$. We now proceed to demonstrate that this parameter regime corresponds to the period-doubling DTC and dynamical freezing (DF) regimes.\\
\begin{figure}[t]
    \centering
    \includegraphics[width=0.47\textwidth]{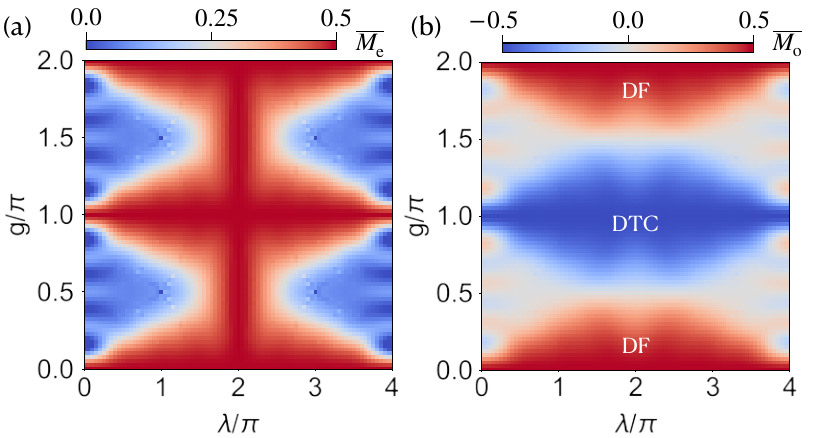}
\caption{{\bf Dynamical freezing and period-doubling DTCs:} (a) Density plot of the time-averaged magnetization at $t=2nT$, $\overline{M_{\rm e}} =\frac{1}{500} \sum_{n=1}^{500} M(2nT)$, when $g_s=g_c=g$. The system exhibits a perfect revival at $t=2nT$ for $g= j \pi \,\ \forall \lambda$ and $\lambda = 2j \pi \,\,\forall g$, where $j \in \mathbb{Z}$. Apart from these points, the system exhibits non-ergodic dynamics for a large parameter regime. (b) Density plot of the time-averaged magnetization at $t=(2n+1)T$, $\overline{M_{\rm o}} =\frac{1}{500} \sum_{n=0}^{499} M\big((2n+1)T\big)$ shows that there is a wide parameter regime where the system exhibits DTC ($\overline{M_{\rm o}} \approx - 0.5$) and DF ($\overline{M_{\rm o}} \approx 0.5$) behaviors. The results presented here are for $N_{\rm sat} = 1901$. }
    \label{fig:period-doubling}
\end{figure}
{\it Period-doubling DTC:} We first analyze the period-doubling DTC that emerges in this system. It is well-known that this model exhibits a period-doubling DTC phase when $g \approx \pi$~\cite{pal2018temporal,kumar2024hilbert}; these oscillations persist indefinitely when $g=\pi$. However, our results for $\zeta^2$ discussed above indicate that the DTC phase may persist over a much wider parameter regime. We explore this issue further by computing the time-averaged magnetization at times $2nT:\,\,  \overline{M}_{\rm e}  =\frac{1}{N} \sum_{n=1}^{N} M(2nT)$, (where $M = \sum_{i=1}^{N_{\rm sat}} S_i^{x}$) for the ${\rm x}$-polarized initial state, $\vert \psi (t=0) \rangle = \left(\prod_{i=1}^{N_{\rm sat}}\vert {\rm +x} \rangle_i \right) \vert {\rm +x} \rangle_c$, where $\vert \pm {\rm x} \rangle$ are the eigenstates of $S_x$ with eigenvalues of $\pm 1/2$. As shown in Fig.~\ref{fig:period-doubling}(a), there is a large parameter regime where  $\overline{M_{\rm e}} \approx 1/2$, indicating initial state memory retention. Intriguingly, when $\lambda = 2\pi$, $\overline{M_{\rm e}} = 1/2~ \forall g$. To elucidate the origin of this non-ergodic behavior, we note that the Floquet unitary can be written as $U_F= U(T,0) = U_0 U_d$, where $U_0 = \exp[-i H_0 T/2] $ and $U_d = \exp[-i H_d T/2]$. When $\lambda = 2\pi$, $S_j^z$ anti-commutes with $U_0$, leading to the following time-evolution operator after 2 periods:
\begin{equation}
U(2T,0) = U_F^2 = U_0 \exp[i g S_c^z] U_0 \exp[i g S_c^z].
\label{eq:MBecho}
\end{equation}
Furthermore, $S_c^z$ commutes (anti-commutes) with $U_0$ when the number of satellite spins is even (odd). Thus, for an odd (even) number of satellite spins, $U(2T,0) = \mathbb{I}$ ($U(2T,0) = \exp[i 2 g S_c^z]$), where $\mathbb{I}$ denotes the Identity matrix. Consequently, the satellite spin magnetization exhibits a perfect revival at $2T$, for all values of $N_{\rm sat}$, thereby signifying an interaction-induced echo. We note that such exact revivals are remarkable, since TTSB usually cannot persist indefinitely in finite-size systems~\cite{huang2024long}. The central spin magnetization on the other hand exhibits an exact revival (sinusoidal oscillations), when $N_{\rm sat}$ is even (odd). We note that this analysis and consequently our conclusions remain unchanged, even when $g_s \ne g_c$. This additional stabilization mechanism leads to the persistence of the DTC phase over a much wider parameter regime than has been reported in Ref.~\cite{kumar2024hilbert}, where the same protocol was studied. \\
Finally, we note that a perfect revival of the satellite-spin magnetization at $2nT$ can be associated with two kinds of non-ergodic behavior: (i) a period-doubling DTC~\cite{khemani2016phase,else2016floquet,yao2017discrete}, and (ii) DF, where the system returns to its initial state at $nT$~\cite{das2010exotic,bhattacharyya2012transverse,haldar2021dynamical,banerjee2024exact,gangopadhay2024counterdiabatic,guo2024dynamical,mukherjee2024floquet}. We distinguish these two kinds of non-ergodic behavior by computing the time-averaged magnetization at $(2n+1)T:\,\, \overline{M}_{\rm o}=\frac{1}{N} \sum_{n=0}^{N-1} M\bigg((2n+1)T\bigg)$. The DTC (DF) phase is characterized by $\overline{M}_{\rm o} \approx -0.5~~(\overline{M}_{\rm o} \approx 0.5)$. Figure~\ref{fig:period-doubling}(b) shows that the system hosts both DTC and DF phases. We gain further insights into the symmetry of this phase diagram by examining the dynamics of the system at $\lambda = 2\pi$. In this case, the time evolution from $0$ to $T/2$ leads to a negative value of $M$  when $\pi/2 < g < 3 \pi/2$. The interaction-induced echo leads to perfect revival of the initial state at $t=2T$. Consequently, $\overline{M}_{\rm o}$ is negative in this regime. An analogous analysis yields that $\overline{M}_{\rm o}$ is positive, when $0 \le g < \pi/2$ and $3 \pi/2 < g \le 2 \pi $. We note that $\overline{M}_{\rm o} = 0$, when $g = \pi/2$ and $3 \pi/2$, marking the transition between the two regimes. A more detailed description of the time evolution of the system is provided in the Supplemental Material~\cite{suppmat}.

\begin{figure}[t]
\centering
    \includegraphics[width=0.45\textwidth]{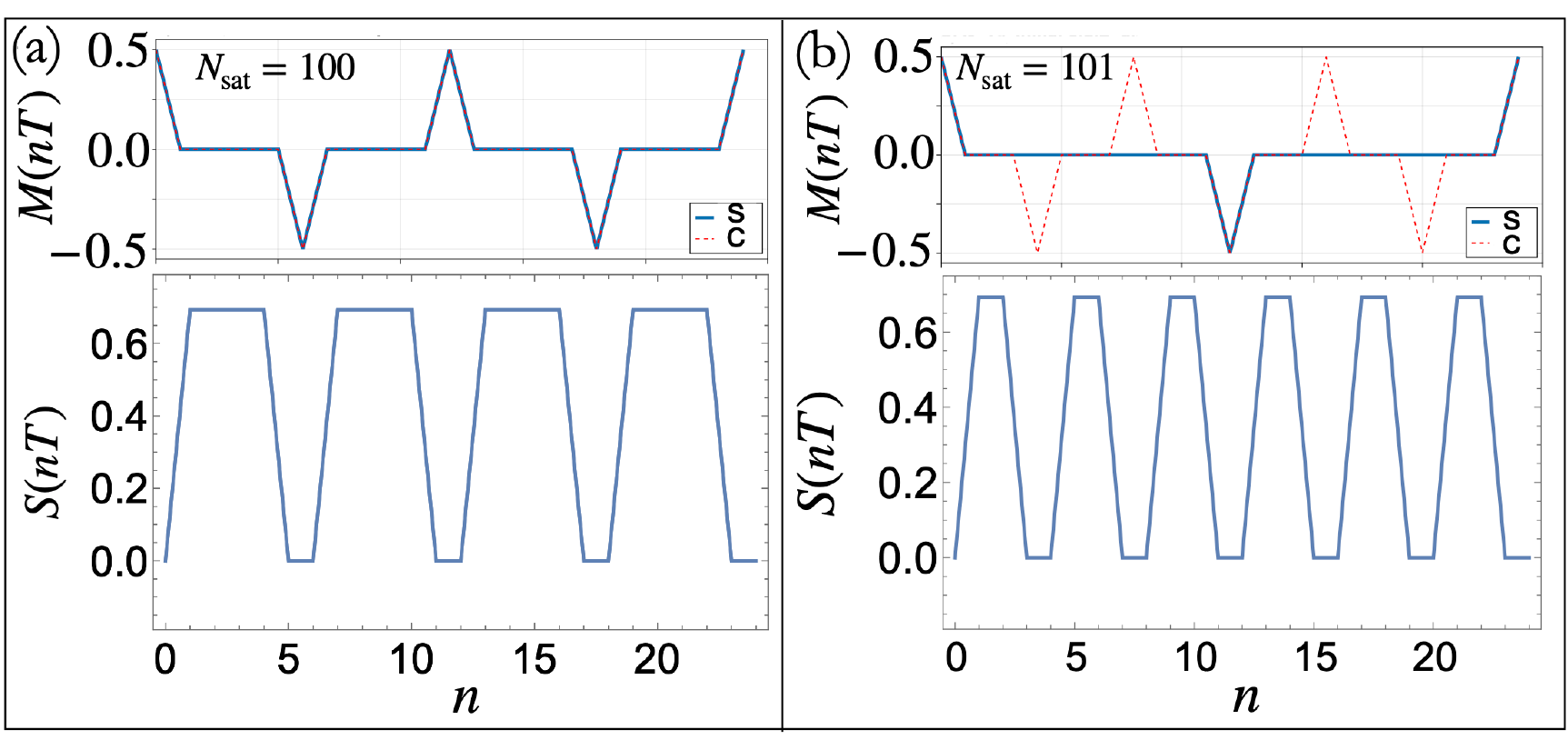}
    \caption{{\bf Clifford-induced dynamics:} (a)-(b) The time-evolution of the magnetization and entanglement entropy at the Clifford points, $\lambda = \pi$ and $g=\pi/2$ for $N_{\rm sat}=100$ and $N_{\rm sat}=101$. Both the satellite and central spin magnetizations oscillate with a period $12 T $ for even $N_{\rm sat}$. However, for odd $N_{\rm sat}$, the satellite and central spin magnetizations oscillate with a period of $24 T$ and $8T$ respectively. This is accompanied by entanglement entropy oscillations with a period of $6T$ and $4T$ for even and odd $N_{\rm sat}$ respectively. The dynamics naturally generate Bell-cat states. Additionally, metrologically valuable spin-cat states emerge for odd $N_{\rm sat}$.}
    \label{fig:DTC-2}
\end{figure}

\emph{Higher-period oscillations and entanglement steering:} We now proceed to investigate whether this model can host exact oscillations with a period longer than $2T$. A natural starting point is to identify parameter points where $U_0$ and $U_d$ act as Clifford gates~\cite{gottesman1997stabilizer,gottesman1998heisenberg}. In that case, the Clifford group structure can lead to exact periodic oscillations for any initial stabilizer state~\cite{aaronson2004improved,sommers2023crystalline}. In our model, these points correspond to $\lambda = (2m+1) \pi$ and $g= (2n+1) \pi/2$, $\forall \, m,n \in \mathbb{Z}$. For simplicity, we focus on $\lambda = \pi$ and $g=\pi/2$ in our analysis, and examine the time-evolution of the fully x-polarized initial state, $\vert \psi (t=0) \rangle = \left(\prod_{i=1}^{N_{\rm sat}}\vert {\rm +x} \rangle_i \right) \vert {\rm +x} \rangle_c$. As shown in Fig.~\ref{fig:DTC-2}(a)-(b), the central and satellite spin magnetizations oscillate with a period of $8T\,\, (12T)$ and $24T\,\, (12T)$ respectively, when $N_{\rm sat}$ is odd (even). These exact revivals result from an interaction-induced echo, and the difference between odd and even $N_{\rm sat}$ originates from the parity-dependence of the phases accumulated during the time-evolution. Further details about the dynamics including the states generated and the stabilizer evolution, are provided in the Supplemental Material~\cite{suppmat}.

We now examine the states generated during the system's time evolution. Intriguingly, we find that our protocol naturally leads to the generation of Bell-cat states~\cite{vlastakis2015characterizing,lajci2024topologically}. We note that generating such maximally entangled states typically requires dedicated state-preparation protocols. However, in this model, they are naturally generated by the dynamics.

We first analyze the time-evolution of this system when $N_{\rm sat}$ is even. The system evolves to the Bell-cat state $\vert \phi^{\rm (BC)} \rangle$ at $t=(12 q+3)T$ for $q \in \mathbb{Z}, \, q \ge 0$:
\begin{equation}
    \vert \phi^{\rm (BC)}_{\pm} \rangle = \frac{1}{\sqrt{2}} \left(\prod_{j=1}^{N_{\rm sat}}\vert {\rm +x} \rangle_j \right)\vert {\rm \pm x} \rangle_c+ \alpha \left(\prod_{j=1}^{N_{\rm sat}}\vert {\rm -x} \rangle_j \right)\vert {\rm \mp x} \rangle_c ,
    \label{eq:xxCat}
\end{equation}
where $\vert \phi^{\rm (BC)}_{+} \rangle$ ($\vert \phi^{\rm (BC)}_{-} \rangle$) with $\alpha = i\, (-i) $ corresponds to $N_{\rm sat} = 4j \,\, (4j+2)$ for $j \in \mathbb{Z}$. Subsequently, in both of these cases, the system evolves to the state $\left(\prod_{j=1}^{N_{\rm sat}}\vert {\rm -x} \rangle_i \right)\vert {\rm -x} \rangle_c $ at $t=(12 q+6)T$, and the system returns to its initial state, $\left(\prod_{j=1}^{N_{\rm sat}}\vert {\rm +x} \rangle_j \right)\vert {\rm +x} \rangle_c$ at the end of every 12 drive periods. This leads to oscillations of the magnetization with a period of $12T$. 

The time-evolution of the system when $N_{\rm sat}$ is odd is more intricate, and a detailed discussion is provided in Ref.~\cite{suppmat}; here we discuss some important aspects of these dynamics. Unlike the even $N_{\rm sat}$ case however, here the central and satellite spin magnetizations oscillate with different periods. We start by noting that at $t= (24 q+4)T$, the system evolves to the state:
\begin{equation}
     \vert \psi^{\rm(sc)} \rangle = \dfrac{1}{\sqrt{2}}\Bigg[\left(\prod_{j=1}^{N_{\rm sat}}|{\rm +z}\rangle_j\right)-\left(\prod_{j=1}^{N_{\rm sat}}|{\rm -z}\rangle_j\right) \Bigg] |{\rm -x}\rangle_c
\end{equation}
where $\vert \pm {\rm z} \rangle$ represent the eigenstates of $\sigma_z$, corresponding to the eigenvalues of $\pm 1$. This leads to oscillations of the central spin magnetization with a period of $8T$. Interestingly, $ \vert \psi^{\rm(sc)} \rangle$ can be useful as a resource for magnetic field sensing, since the satellite spins are in a `spin-cat' state~\cite{dooley2013collapse,huang2018achieving}. Moreover, after $t=(24q+6)T$, the system evolves to the maximally entangled $\vert \phi^{\rm (BC)}_{+} \rangle$ state (defined in Eq.~\ref{eq:xxCat}) with $\alpha =-i$. This in turn leads to the state $\left(\prod_{j=1}^{N_{\rm sat}}\vert {\rm -x} \rangle_j \right)\vert {\rm -x} \rangle_c $ at $t=(24q+12)T$; consequently, the satellite spin magnetization oscillates with a period of $24T$. It is clear from this analysis that the Clifford-induced dynamics lead to entanglement oscillations. To completely characterize the nature of these oscillations, we compute the entanglement between the central and the satellite spins, $S=-\rho_c \ln(\rho_c)$, where $\rho_c$ is the reduced density matrix of the central spin. Our analysis shown in Fig.~\ref{fig:DTC-2}(b) shows that our protocol steers the system through an entangled trajectory, where $S$ oscillates with a period of $4T$ and $6T$ for odd and even values of $N_{\rm sat}$, respectively.

We note that while our discussion has focused on the quantum resonant dynamics at $\lambda = \pi$ and $g = \pi/2$, the spectral response of this system away from the Clifford points also exhibits a parity dependence. The Fourier peaks remain pinned for small detunings away from the Clifford points for even $N_{\rm sat}$, while they broaden and shift for odd $N_{\rm sat}$ over the finite time window examined here (see Supplemental Material~\cite{suppmat}). Our results thus provide a mechanism to engineer entanglement oscillations with parity-dependent periods in collective spin systems.

\begin{figure}[t]
    \includegraphics[width=0.47\textwidth]{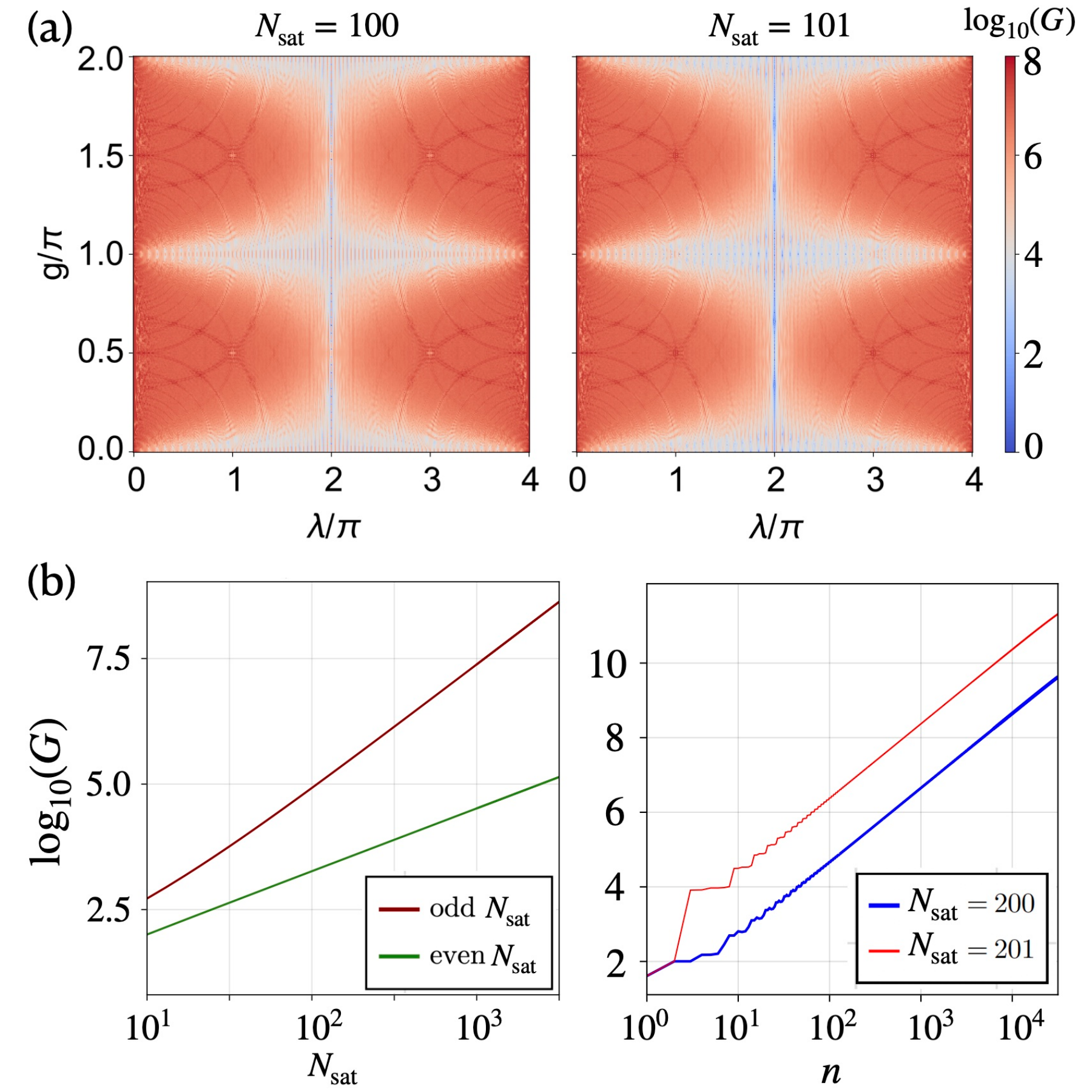}
    \caption{{\bf Multiparameter quantum metrology:} (a) The density plot of $G$ (Eq.~\ref{eq:QFI}), when $N_{\rm sat} =100$ and $101$ at $t=100T$. Both the Clifford and chaotic regimes are associated with a high value of $G$. (b) The scaling of $G$ with $N_{\rm sat}$ (left panel) at $t=100T$ and the number of drive periods, $n$ (right panel) at the Clifford point, $\lambda = \pi$ and $g = \pi/2$. The QFI, $G$, scales as $G \propto n^2 N_{\rm sat}^2$ and  $G \propto n^2 N_{\rm sat}$ for odd and even $N_{\rm sat}$ respectively. The Heisenberg-limited sensitivity for odd $N_{\rm sat}$ originates from the emergence of spin-cat states at $t=(24q+4)T$ for $q\in \mathbb{Z}$.}
    \label{fig:QFI}
\end{figure}
\emph{Multiparameter quantum metrology:} We have demonstrated that the Floquet CSM exhibits entanglement oscillations at Clifford points. We now demonstrate that the states generated during these oscillations can be employed for quantum-enhanced sensing of the interaction strength, $\lambda$ and the magnetic field, $g$. To characterize this sensitivity, we note that the uncertainty in the joint measurement of $\lambda$ and $g$ can be obtained from the $2 \times 2$ quantum Fisher information (QFI) matrix, $\mathcal{F}$, whose elements are given by:
$\mathcal{F}_{\lambda g}= 
4\,\mathrm{Re}
\left[
\langle \partial_\lambda \psi \vert \partial_g \psi \rangle -\langle \partial_\lambda \psi \vert \psi \rangle \langle \psi \vert \partial_g \psi \rangle
\right]$~\cite{liu2020quantum,demkowicz2020multi,mondal2025multicritical}. The equally-weighted uncertainty in measuring $\lambda$ and $g$ is given by $\delta \lambda^2 +\delta g^2 \ge G^{-1}$, where
\begin{equation}
G = \frac{\mathcal{F}_{\lambda\lambda} + \mathcal{F}_{gg}}{\mathcal{F}_{\lambda\lambda}\mathcal{F}_{gg} - \mathcal{F}_{\lambda g}\mathcal{F}_{g \lambda }}.
\label{eq:QFI}
\end{equation}
$G$ characterizes genuine multipartite entanglement~\cite{liu2020quantum}, and it is enhanced (suppressed) in the Clifford and chaotic (period-doubling DTC and DF) regimes (Fig.~\ref{fig:QFI}(a)). Interestingly, the qualitative similarity of Fig.~\ref{fig:QFI}(a) and Fig.~\ref{fig:schematic}(b) originates from the fact that both $G$ and the period-blind order parameter $\zeta^2$ exhibit similar temporal growth up to the Ehrenfest time, $t_E$ ~\cite{sinha2024classical,lerose2020bridging}; after $t_E$ however, $\zeta^2$ saturates~\cite{sinha2024classical,vivek2025self}, while $G$ continues to grow~\cite{fiderer2018quantum}. Thus, unlike $\zeta^2$, $G$ determines the metrological sensitivity of the system at all times.\\
In order to characterize the scaling behavior of $G$, we have taken a scaling form for $G (t=nT, N_{\rm sat}) \propto n^{\alpha} N_{\rm sat}^{\beta}$. For classical probes, the best possible sensitivity corresponds to $\beta = 1$; this is known as the standard quantum limit (SQL). By employing quantum resources, it is possible to surpass the SQL and reach the Heisenberg limit (HL) corresponding to $\alpha=\beta=2$~\cite{montenegro2024quantum}. \\
As shown in Fig.~\ref{fig:QFI}(b), at the Clifford points, the scaling of $G$ always surpasses the SQL. Interestingly, the parity dependence of the states generated during the oscillations is reflected in the exponents $\alpha$ and $\beta$. In particular, for odd $N_{\rm sat}$, the dynamics generate metrologically valuable spin-cat states in addition to the Bell-cat states, thereby enabling HL sensitivity ($\alpha=\beta=2$). In contrast, the Bell-cat states generated for even $N_{\rm sat}$, lead to a QFI scaling with $\alpha=2$ and $\beta=1$. The chaotic regime does not exhibit a clear scaling behavior~\cite{suppmat}. Our results reveal that Clifford-induced entanglement in the driven CSM can enable quantum-enhanced multiparameter sensing. 

\emph{Conclusion and Outlook:} In this work, we introduced interaction-induced echo protocols to realize exact quantum revivals and discrete time crystals in the driven CSM. A key result is that tuning the interaction strength to $\lambda = 2  \pi$ yields exact period-doubling oscillations for arbitrary $N_{\rm sat}$, giving rise to both DTCs and dynamical freezing regimes. Furthermore, at $\lambda = \pi$  and $g = \pi/2$, the Clifford group structure is induced in the Floquet dynamics, leading to exact higher-period oscillations that dynamically generate Bell-cat and spin-cat states. These entanglement oscillations are characterized by a parity-dependent period. Finally, we have demonstrated that the Clifford-induced entanglement dynamics can be employed for multiparameter quantum metrology, with odd $N_{\rm sat}$ achieving the Heisenberg limit. We conclude that the driven CSM can exhibit a rich array of non-equilibrium phenomena, that can be realized in near-term experiments with NMR~\cite{pal2018temporal} and NV centers~\cite{geng2021ancilla}, where period-doubling DTCs have already been observed.\\
There are several natural extensions of this work. Firstly, it would be interesting to investigate the non-equilibrium phases that would emerge in driven central spin-$S$ models with $S>1/2$. It would also be intriguing to examine the nature of the entanglement dynamics when interactions are introduced between the satellite spins. Finally, the dynamics of central spin systems subjected to aperiodic driving would be a fruitful avenue for future research.
\section*{acknowledgments} 
SC thanks DST, India for support through SERB project SRG/2023/002730 and DST/FFT/NQM/QSM/2024/3.

\bibliographystyle{apsrev4-2}
\bibliography{ref}

\end{document}


\title{Supplemental Material: The Floquet central spin model: A platform to realize time crystals, entanglement steering, and multiparameter metrology}
\author{Hillol Biswas}
\author{Sayan Choudhury}
\email{sayanchoudhury@hri.res.in}
\affiliation{Harish-Chandra Research Institute, Chhatnag Road, Jhunsi, Allahabad 211019, India}
\affiliation{Homi Bhabha National Institute, Training School Complex, Anushakti Nagar, Mumbai 400094, India}
\maketitle
In this Supplemental Material, we provide (i) an alternative analysis of the interaction-induced echo protocol at $\lambda = 2 \pi$, numerical results of the phase diagram for various system sizes, and a discussion of the absence of multifractality in the DTC phase, and (ii) a detailed characterization of the states generated during the higher-period oscillations at the Clifford points, and some additional numerical results on the multiparameter sensing protocol.

\section{An alternative analysis of the interaction-induced echo protocol at $\lambda = 2 \pi$}
\begin{figure*}[t]
    \includegraphics[width=0.95\textwidth]{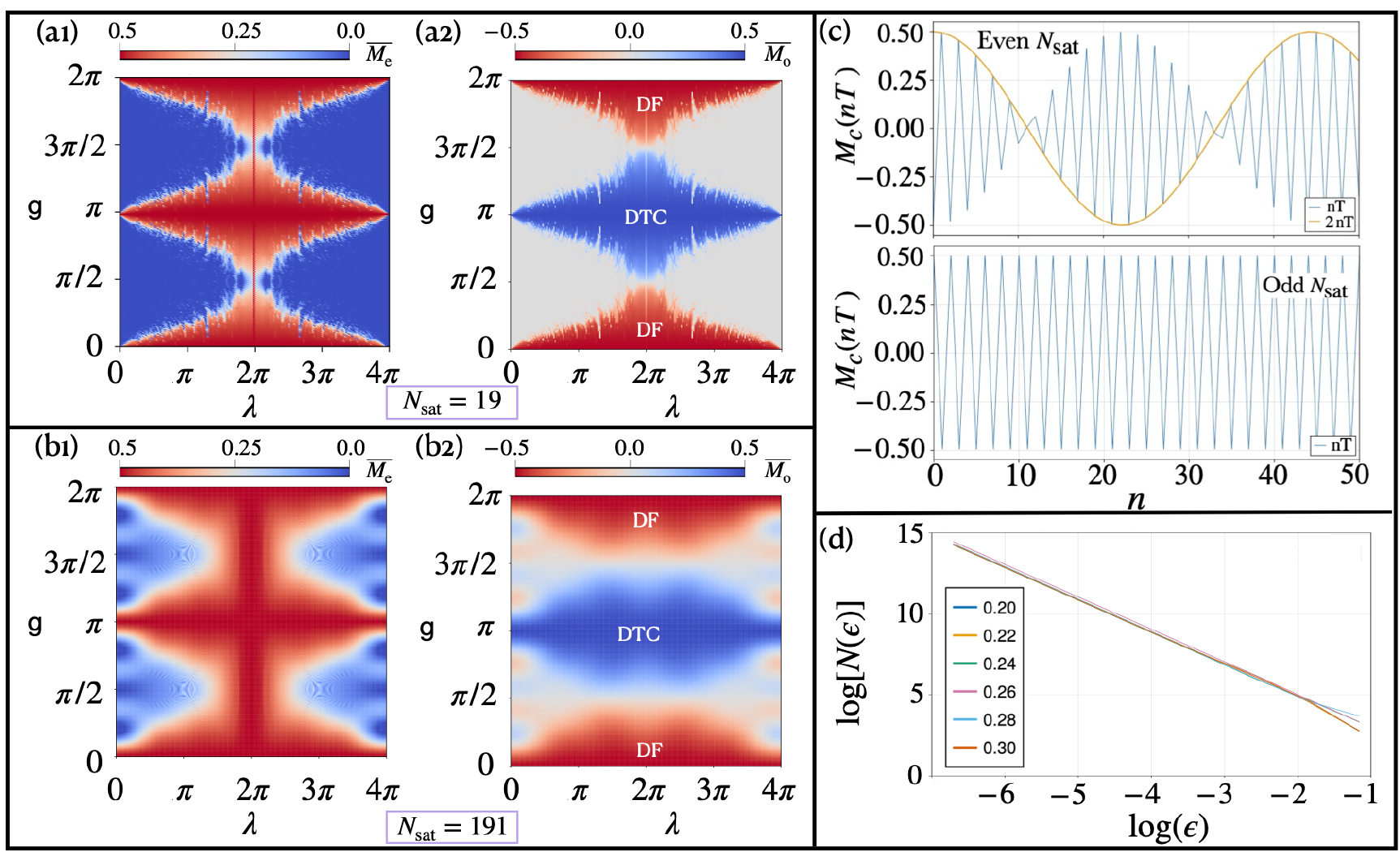}
    \caption{(a1)-(a2) and (b1)-(b2) show the time-averaged magnetization at $t=2nT$ ($\overline{M_{\rm e}}$) and $t=(2n+1)T$ ($\overline{M_{\rm o}}$) for $N_{\rm sat} = 19$ and $N_{\rm sat} = 191$ respectively. While finite-size effects are present for $N_{\rm sat} = 19$, there is no appreciable difference between $N_{\rm sat} = 191$ results and the $N_{\rm sat} = 1901$ results reported in Fig.~2 of the main text. (c) The interaction-induced echo leads to period-doubling oscillations of the central spin magnetization, $M_c(nT)$, only for odd $N_{\rm sat}$. (d) The Minkowski-Bouligand dimension, $d_{\rm MB}$ (Eq.~\ref{eq:fractal}), for the period-doubling DTC is $2$, indicating an absence of fractality.}
    \label{fig:ExtendedPD}
\end{figure*}

In the main text, we have demonstrated that an interaction-induced echo leads to perfect period-doubling oscillations of the satellite spin magnetization, when $\lambda = 2 \pi$ for all values of $g_s$. This in turn leads to a period-doubling DTC for a wide parameter regime, even for small values of $N_{\rm sat}$ (see Fig.~\ref{fig:ExtendedPD} (a1)-(a2) and (b1)-(b2)). In this section, we provide an alternative derivation of this phenomenon. In particular, we examine the dynamics of a fully x-polarized initial state, $\vert \psi (t=0) \rangle = \left(\prod_{i=1}^{N_{\rm sat}}\vert {\rm +x} \rangle_i \right) \vert {\rm +x} \rangle_c$. In the following analysis, all the states are defined modulo an overall phase. After the first drive period, the system evolves to the state:
\begin{eqnarray}
 \vert \psi(t=T) \rangle &=& U_0 \left[\left(\prod_{j=1}^{N_{\rm sat}}(\cos(g_s/2) \vert {\rm +x} \rangle_j - i \sin(g_s/2) \vert {\rm -x} \rangle_j) \right) (\cos(g_c/2) \vert {\rm +x} \rangle_c - i \sin(g_c/2) \vert {\rm -x} \rangle_c)\right] \nonumber \\
  &=& \left[\left(\prod_{j=1}^{N_{\rm sat}}(\cos(g_s/2) \vert {\rm +x} \rangle_j + i \sin(g_s/2) \vert {\rm -x} \rangle_j) \right) (\cos(g_c/2) \vert {\rm +x} \rangle_c - i (-1)^{N_{\rm sat}} \sin(g_c/2) \vert {\rm -x} \rangle_c)\right]
\end{eqnarray}
Consequently, when $N_{\rm sat}$ is odd, we obtain the following: 
\begin{eqnarray}
     \vert \psi(t=2T) \rangle &=& U_0 U_d \left[ \left(\prod_{j=1}^{N_{\rm sat}}(\cos(g_s/2) \vert {\rm +x} \rangle_j + i \sin(g_s/2) \vert {\rm -x} \rangle_j) \right) (\cos(g_c/2) \vert {\rm +x} \rangle_c + i \sin(g_c/2) \vert {\rm -x} \rangle_c)\right]  \nonumber \\
     &=& U_0 \left[\left(\prod_{i=1}^{N_{\rm sat}}\vert {\rm +x} \rangle_i \right) \vert {\rm +x} \rangle_c\right] = \vert \psi(t=0) \rangle.
\end{eqnarray}
Thus, both the satellite spin and central spin magnetization show exact period-doubling oscillations. \\

This situation changes when $N_{\rm sat}$ is even:
\begin{eqnarray}
     \vert \psi(t=2T) \rangle &=& U_0 U_d \left[ \left(\prod_{j=1}^{N_{\rm sat}}(\cos(g_s/2) \vert {\rm +x} \rangle_j + i \sin(g_s/2) \vert {\rm -x} \rangle_j) \right) (\cos(g_c/2) \vert {\rm +x} \rangle_c - i \sin(g_c/2) \vert {\rm -x} \rangle_c)\right]  \nonumber \\
     &=& U_0 \left[\left(\prod_{i=1}^{N_{\rm sat}}\vert {\rm +x} \rangle_i \right) (\cos(g_c) \vert {\rm +x} \rangle_c - i \sin(g_c) \vert {\rm -x} \rangle_c)\right] =  \left(\prod_{i=1}^{N_{\rm sat}}\vert {\rm +x} \rangle_i \right) (\cos(g_c) \vert {\rm +x} \rangle_c - i \sin(g_c) \vert {\rm -x} \rangle_c) \nonumber\\
\end{eqnarray}

Thus, in this case, the satellite spin magnetization exhibits perfect period-doubling oscillations, while the central spin magnetization oscillates sinusoidally. These results are shown in Fig.~\ref{fig:ExtendedPD}(c).\\

Before proceeding to examine the dynamics at the Clifford points, it is worth noting that the boundaries between the chaotic and the DTC phases in the periodically driven infinite-range Ising model exhibit a self-similar fractal structure~\cite{giachetti2023fractal}. A natural question then is whether the period-doubling DTC studied in this work exhibits a similar fractal structure. We have performed this analysis by calculating the fractal dimension, also known as the Minkowski-Bouligand dimension, $d_{\rm MB}$. To this end, we uniformly cover the period-blind phase diagram (Fig.~1(b) of the main text) with a square grid of side $\epsilon$ and count the number of such boxes ($N(\epsilon)$) lying on the boundary between the DTC and the chaotic phase. The fractal dimension, $d_{\rm MB}$ is given by
\begin{equation}
    d_{\rm MB}=-\lim_{\epsilon \to 0}\frac{\log N(\epsilon)}{\log \epsilon}
    \label{eq:fractal}
\end{equation}
An integer value of $d_{\rm MB}$ indicates a smooth manifold, whereas a fractional value indicates fractal behavior. Figure~\ref{fig:ExtendedPD}(d) demonstrates that the fractal dimension $d_{\rm MB}=2$ in both cases, indicating the absence of fractality.

\section{Detailed analysis of the entanglement oscillations at the Clifford Points} 

In this section, we analyze the dynamics at the Clifford points in greater detail and elucidate the nature of the states generated. This analysis demonstrates that the exact revivals at these points, which originate from the Clifford group structure, can alternatively be understood from the perspective of interaction-induced echoes.

\subsection{Even $N_{\rm sat}$}

We begin our analysis by examining the states generated during the time-evolution of the system, when $\lambda = \pi$ and $g=\pi/2$, for even values of $N_{\rm sat}$. We start by noting that for a single spin:
\begin{equation}
    U_d=\exp{(i\dfrac{\pi}{2}S_z)}=\dfrac{1}{\sqrt{2}}\begin{pmatrix}
    1+i&0\\
    0&1-i
\end{pmatrix}.
\end{equation} 
We first analyze the evolution of one satellite spin for one drive period and then extrapolate the results for $N_{\rm sat}$ spins. In the following analysis, all the states are defined modulo an overall phase. Furthermore, we employ the following definitions for the $|\pm \rm x\rangle$,  $|\pm \rm y\rangle$, and  $|\pm \rm z\rangle$ states:\\
\begin{widetext}
\begin{equation}
   |\pm \rm x\rangle = \dfrac{1}{\sqrt{2}}\begin{bmatrix}
    \pm 1 \\ 1
\end{bmatrix}, \, |\pm \rm y\rangle = \dfrac{1}{\sqrt{2}}\begin{bmatrix}
    \mp i \\ 1
\end{bmatrix}, \, |\pm z\rangle = \begin{bmatrix}
    1 \\ 0
\end{bmatrix}, \, {\rm and} \, |-z\rangle = \begin{bmatrix}
    0 \\ 1
\end{bmatrix}
\end{equation}
\end{widetext}

The x-polarized initial state, $|\phi (t=0)\rangle=|{\rm +x}\rangle_s |{\rm +x}\rangle_c$, evolves after one drive period to:
\begin{equation}
    \vert \phi(t=T) \rangle = \frac{1}{\sqrt{2}}\bigg(|{\rm +z}\rangle_s|{\rm +x}\rangle_c-|{\rm -z}\rangle_s|{\rm -x}\rangle_c \bigg)
\end{equation}
Generalizing this analysis to the case of $N_{\rm sat}$ satellite spins, when $\vert \psi (t=0) \rangle = \left(\prod_{i=1}^{N_{\rm sat}}\vert {\rm +x} \rangle_i \right) \vert {\rm +x} \rangle_c$, we obtain:

\begin{widetext}

\begin{equation}
    \vert \psi(t=T) \rangle = \frac{1}{\sqrt{2}}\bigg[\left(\prod_{j=1}^{N_{\rm sat}}|{\rm +z}\rangle_j\right)|{\rm +x}\rangle_c+ (-i)^{N_{\rm sat}+1} \left(\prod_{j=1}^{N_{\rm sat}}|{\rm -z}\rangle_j\right)|{\rm -x}\rangle_c\bigg]
    \label{eq:kick1}
\end{equation}
\end{widetext}
The action of $U_d$ on this state is straightforward, leading to the following state at $t=3T/2$:

\begin{widetext}

\begin{equation}
    \vert \psi(t=3T/2) \rangle = \frac{1}{\sqrt{2}}\Bigg[\left(\prod_{j=1}^{N_{\rm sat}}|{\rm +z}\rangle_j\right)|{\rm +y}\rangle_c -i \left(\prod_{j=1}^{N_{\rm sat}}|{\rm -z}\rangle_j\right)|{\rm -y}\rangle_c\Bigg]
    \label{1st_kick}
\end{equation}
\end{widetext}
The evolution of the system from $t=3T/2$ to $t=2T$ can be split into 4 classes depending on the value of $N_{\rm sat}$. For even $N_{\rm sat}$, we obtain the following states at $t=2T$:\\
\begin{widetext}

\begin{align}
    &N_{\rm sat}=4n : \vert \psi(t=2T) \rangle = \dfrac{1}{\sqrt{2}}\Bigg[\left (\prod_{j=1}^{N_{\rm sat}}|{\rm +y}\rangle_j\right)|{\rm -y}\rangle_c -i\left(\prod_{j=1}^{N_{\rm sat}}|{\rm -y}\rangle_j\right)|{\rm +y}\rangle_c \Bigg], \nonumber\\
    &N_{\rm sat}=4n+2 : \vert \psi(t=2T) \rangle = \dfrac{1}{\sqrt{2}}\Bigg[\left(\prod_{j=1}^{N_{\rm sat}}|{\rm +y}\rangle_j\right)|{\rm +y}\rangle_c+ i \left(\prod_{j=1}^{N_{\rm sat}}|{\rm -y}\rangle_j\right)|{\rm -y}\rangle_c\Bigg].
    \label{classified-even}
\end{align}
\end{widetext}
where $n$ is a positive integer. Over the course of the next drive period, the system evolves to the following Bell-cat states:
\begin{widetext}
\begin{align}
    &N_{\rm sat}=4n: \vert \psi(t=3T) \rangle = \dfrac{1}{\sqrt{2}}\Bigg[\left(\prod_{j=1}^{N_{\rm sat}}|{\rm +x}\rangle_j\right)|{\rm +x}\rangle_c + i \left(\prod_{j=1}^{N_{\rm sat}}|{\rm -x}\rangle_j\right)|{\rm -x}\rangle_c \Bigg] \nonumber\\
    &N_{\rm sat}=4n+2: \vert \psi(t=3T) \rangle = \dfrac{1}{\sqrt{2}}\Bigg[\left(\prod_{j=1}^{N_{\rm sat}}|{\rm +x}\rangle_j\right)|{\rm -x}\rangle_c - i \left(\prod_{j=1}^{N_{\rm sat}}|{\rm -x}\rangle_j\right)|{\rm +x}\rangle_c \Bigg].
\end{align}
\end{widetext}
We conclude that the application of $U_F^3$ transforms the initial `+x'-polarized state to a Bell-cat state. A further evolution under $U_F^3$ creates a `-x'-polarized state, and the system returns to its initial state after 12 drive periods. \\

\begin{table}
\centering
\caption{Time-evolution of the stabilizer generators $\mathcal{G}_c$ and $\mathcal{G}_s^{(1)}$ over 6 drive periods for $N_{\rm sat} = 4n$ and $N_{\rm sat} = 4n+2$}
\renewcommand{\arraystretch}{2.2}
\begin{tabular}{|c|c|c|c|c|}
\hline
${\rm Drive \, \, Period}$ & $\mathcal{G}_c (N_{\rm sat} = 4n) $ & $\mathcal{G}_s^{(1)} (N_{\rm sat} = 4n)$ & $\mathcal{G}_c (N_{\rm sat} = 4n+2)$ & $\mathcal{G}_s^{(1)} (N_{\rm sat} = 4n+2)$ \\
\hline
0  & $+\sigma_c^x$ & $+\sigma_1^x$ & $+\sigma_c^x$ & $+\sigma_1^x$\\
\hline
1  & $-\sigma_c^y \prod_{i=1}^{N_{\rm sat}}\sigma_i^x$ & $+\sigma_c^x\,\sigma_1^z$ & $+\sigma_c^y \prod_{i=1}^{N_{\rm sat}}\sigma_i^x$ & $+\sigma_c^x\,\sigma_1^z$\\
\hline
2  & $-\sigma_c^x \prod_{i=1}^{N_{\rm sat}}\sigma_i^z$ & $-\sigma_c^z\,\sigma_1^z\,\prod_{i=2}^{N_{\rm sat}}\sigma_i^x$ & $+\sigma_c^x \prod_{i=1}^{N_{\rm sat}} \sigma_i^z$ & $+\sigma_c^z\,\sigma_1^z\,\prod_{i=2}^{N_{\rm sat}}\sigma_i^x$\\
\hline
3  & $+\sigma_c^y \prod_{i=1}^{N_{\rm sat}} \sigma_i^z$ & $+\sigma_c^z\,\sigma_1^z\,\prod_{i=2}^{N_{\rm sat}}\sigma_i^y$ & $-\sigma_c^y \prod_{i=1}^{N_{\rm sat}} \sigma_i^z$ & $ -\sigma_c^z\,\sigma_1^z\,\prod_{i=2}^{N_{\rm sat}} \sigma_i^y$ \\
\hline
4  & $+\sigma_c^x \prod_{i=1}^{N_{\rm sat}} \sigma_i^y$ & $-\sigma_c^y\,\sigma_1^z$ & $-\sigma_c^x \prod_{i=1}^{N_{\rm sat}} \sigma_i^y$ & $-\sigma_c^y\,\sigma_1^z$\\
\hline
5  & $-\sigma_c^y$ & $-\sigma_1^y$ & $-\sigma_c^y$ & $-\sigma_1^y$\\
\hline
6  & $-\sigma_c^x$ & $-\sigma_1^x$ & $-\sigma_c^x$ & $-\sigma_1^x$\\
\hline
\end{tabular}
\label{tab:stabilizer}
\end{table}

Before concluding this discussion, it is worth examining this perfect quantum recurrence from an alternative perspective. As discussed in the main text, $U_0$ and $U_d$ form Clifford gates when $\lambda = \pi$ and $g_s=g_c = \pi/2$, and the x-polarized initial state is a stabilizer state. The finiteness of the stabilizer group then ensures that a perfect revival would occur at long enough times. Furthermore, due to the permutation symmetry of the CSM, the revival time depends only on the parity of $N_{\rm sat}$. To elucidate this, it is instructive to examine the time-evolution of the generators of the stabilizer group. The stabilizer group generators for the x-polarized initial state are $\mathcal{G}_c = \sigma_c^x$ and $\mathcal{G}_s^{(i)} = \sigma_i^x \, \, \forall i \in \{1,N_{\rm sat}\}$. After $m$ drive periods, the stabilizer generator $\mathcal{G}_c (mT) = U_F^m \sigma_c^x U_F^{-m}$ takes the form  $\mathcal{G}_c (mT) = (-1)^a \sigma_c^{\alpha} \prod_{i=1}^{N_{\rm sat}} \sigma_i^{\beta}$. Similarly, the time-evolution of the satellite spin stabilizer generator, $\mathcal{G}_s^{(j)}$ takes the form $\mathcal{G}_s^{(j)} (mT) = (-1)^b \sigma_c^{\gamma} \sigma_j^{\delta} \prod_{i\ne j} \sigma_i^{\epsilon}$. We note that in these expressions, $a, b \in \{0,1\}$ and $\alpha, \beta, \gamma, \delta, \epsilon \in \{0,x,y,z\}$, where $\sigma^{0}$ represents the $2 \times 2$ Identity matrix. Thus, the time-evolution of the stabilizer generators is completely determined by a finite set of parameters for all values of $N_{\rm sat}$. Consequently, the system exhibits a finite-time recurrence for any value of $N_{\rm sat}$. We provide the stabilizer evolution for both the central spin and one satellite spin for $N_{\rm sat}=4n$ and $N_{\rm sat}=4n+2$ in Tab.~\ref{tab:stabilizer} for 6 drive periods. The time-evolution of the stabilizer generators at later times is determined by the following rule:
\begin{equation}
\mathcal{G}\bigg((m+6)T\bigg) = -\,\mathcal{G}(mT), \,\, \mathcal{G}\bigg((m+12)T\bigg) = \mathcal{G}(mT).
\end{equation}

\begin{table}
\centering
\renewcommand{\arraystretch}{2.5}
\caption{ Stabilizer evolution over 12 drive periods for odd $N_{\rm sat}$ for $\lambda = \pi$ and $g=\pi/2$}
\begin{tabular}{|c|c|c|c|c|}
\hline
${\rm Drive \, \, Period}$ & $\mathcal{G}_c$ ($N_{\rm sat}=4n+1$) & $\mathcal{G}_s^{(1)}$ ($N_{\rm sat}=4n+1$) & $\mathcal{G}_c$ ($N_{\rm sat}=4n+3$) & $\mathcal{G}_s^{(1)}$ ($N_{\rm sat}=4n+3$) \\
\hline
0  & $\sigma_c^x$ & $\sigma_1^x$ & $\sigma_c^x$ & $\sigma_1^x$ \\
\hline
1  & $+\sigma_c^z \prod_{i=1}^{N_{\rm sat}} \sigma_i^x$ & $+\sigma_c^x \sigma_1^z$ 
   & $-\sigma_c^z \prod_{i=1}^{N_{\rm sat}} \sigma_i^x$ & $+\sigma_c^x \sigma_1^z$ \\
\hline
2  & $-\sigma_c^z \prod_{i=1}^{N_{\rm sat}} \sigma_i^y$ 
   & $-\sigma_c^y \sigma_1^z \prod_{i=2}^{N_{\rm sat}}\sigma_i^x$
   & $+\sigma_c^z \prod_{i=1}^{N_{\rm sat}} \sigma_i^y$
   & $+\sigma_c^y \sigma_1^z \prod_{i=2}^{N_{\rm sat}}\sigma_i^x$ \\
\hline
3  & $-\sigma_c^y$ 
   & $-\sigma_1^y \prod_{i=2}^{N_{\rm sat}}\sigma_i^z$
   & $-\sigma_c^y$
   & $+\sigma_1^y \prod_{i=2}^{N_{\rm sat}}\sigma_i^z$ \\
\hline
4  & $-\sigma_c^x$
   & $-\sigma_1^x \prod_{i=2}^{N_{\rm sat}}\sigma_i^y$
   & $-\sigma_c^x$
   & $+\sigma_1^x \prod_{i=2}^{N_{\rm sat}}\sigma_i^y$ \\
\hline
5  & $-\sigma_c^z \prod_{i=1}^{N_{\rm sat}} \sigma_i^x$
   & $-\sigma_c^x \sigma_1^z \prod_{i=2}^{N_{\rm sat}}\sigma_i^x$
   & $+\sigma_c^z \prod_{i=1}^{N_{\rm sat}} \sigma_i^x$
   & $+\sigma_c^x \sigma_1^z \prod_{i=2}^{N_{\rm sat}}\sigma_i^x$ \\
\hline
6  & $+\sigma_c^z \prod_{i=1}^{N_{\rm sat}} \sigma_i^y$
   & $+\sigma_c^y \sigma_1^z \prod_{i=2}^{N_{\rm sat}}\sigma_i^y$
   & $-\sigma_c^z \prod_{i=1}^{N_{\rm sat}} \sigma_i^y$
   & $-\sigma_c^y \sigma_1^z \prod_{i=2}^{N_{\rm sat}}\sigma_i^y$ \\
\hline
7  & $+\sigma_c^y$
   & $+\sigma_1^y \prod_{i=2}^{N_{\rm sat}}\sigma_i^x$
   & $+\sigma_c^y$
   & $-\sigma_1^y \prod_{i=2}^{N_{\rm sat}}\sigma_i^x$ \\
\hline
8  & $\sigma_c^x$
   & $+\sigma_1^x \prod_{i=2}^{N_{\rm sat}}\sigma_i^z$
   & $\sigma_c^x$
   & $-\sigma_1^x \prod_{i=2}^{N_{\rm sat}}\sigma_i^z$ \\
\hline
9  & $+\sigma_c^z \prod_{i=1}^{N_{\rm sat}} \sigma_i^x$
   & $+\sigma_c^x \sigma_1^z \prod_{i=2}^{N_{\rm sat}}\sigma_i^y$
   & $-\sigma_c^z \prod_{i=1}^{N_{\rm sat}} \sigma_i^x$
   & $-\sigma_c^x \sigma_1^z \prod_{i=2}^{N_{\rm sat}}\sigma_i^y$ \\
\hline
10 & $-\sigma_c^z \prod_{i=1}^{N_{\rm sat}} \sigma_i^y$
   & $-\sigma_c^y \sigma_1^z$
   & $+\sigma_c^z \prod_{i=1}^{N_{\rm sat}} \sigma_i^y$
   & $-\sigma_c^y \sigma_1^z$ \\
\hline
11 & $-\sigma_c^y$
   & $-\sigma_1^y$
   & $-\sigma_c^y$
   & $-\sigma_1^y$ \\
\hline
12 & $\sigma_c^x$ & $\sigma_1^x$ & $\sigma_c^x$ & $\sigma_1^x$ \\
\hline
\end{tabular}
\label{tab:stabilizer_cycle}
\end{table}

\subsection{Odd $N_{\rm sat}$}

We now proceed to further analyze the time-evolution of the system when $N_{\rm sat}$ is odd, for $\lambda = \pi$ and $g=\pi/2$. We start by noting that the time-evolution of the system up to $t=3T/2$ is similar for all values of $N_{\rm sat}$. The only dependence on $N_{\rm sat}$ comes through a relative phase. However, after that, the dynamics of these higher-period oscillations are qualitatively different for even and odd $N_{\rm sat}$. For odd $N_{\rm sat}$, the system evolves to the following state at $t=2T$:
\begin{align}
    &N_{\rm sat}=4n+1 : \vert \psi(t=2T) \rangle =\dfrac{1}{\sqrt{2}}\Bigg[\left(\prod_{j=1}^{N_{\rm sat}}|{\rm -y}\rangle_j\right)|{\rm -z}\rangle_c +i \left(\prod_{j=1}^{N_{\rm sat}}|{\rm +y}\rangle_j \right)\vert{\rm +z}\rangle_c\Bigg],\\\nonumber
    &N_{\rm sat}=4n+3 : \vert \psi(t=2T) \rangle = \dfrac{1}{\sqrt{2}}\Bigg[\left(\prod_{j=1}^{N_{\rm sat}}|{\rm -y}\rangle_j\right)|{\rm +z}\rangle_c +i \left(\prod_{j=1}^{N_{\rm sat}}|{\rm +y}\rangle_j \right)\vert{\rm -z}\rangle_c\Bigg],
    \label{classified-odd}
\end{align}
The system evolves to the satellite-spin cat state over the course of the next drive period, leading to the following state at $t=3T$:
\begin{align}
    &N_{\rm sat}=4n+1 : \vert \psi(t=3T) \rangle = \dfrac{1}{\sqrt{2}}\Bigg[\left(\prod_{j=1}^{N_{\rm sat}}|{\rm +x}\rangle_j\right)-i\left(\prod_{j=1}^{N_{\rm sat}}|{\rm -x}\rangle_j\right)\Bigg]|{\rm +y}\rangle_c \\ \nonumber
    &N_{\rm sat}=4n+3: \vert \psi(t=3T) \rangle = \dfrac{1}{\sqrt{2}}\Bigg[\left(\prod_{j=1}^{N_{\rm sat}}|{\rm +x}\rangle_j\right) +i\left(\prod_{j=1}^{N_{\rm sat}}|{\rm -x}\rangle_j\right)\Bigg]|{\rm +y}\rangle_c
\end{align}

The application of $U_d$ on this state leads to the following state at $t=3.5T$:
\begin{align}
    &N_{\rm sat}=4n+1 : \vert \psi(t=3.5T) \rangle =\dfrac{1}{\sqrt{2}}\Bigg[\left(\prod_{j=1}^{N_{\rm sat}}|{\rm +y}\rangle_j\right)-i\left(\prod_{j=1}^{N_{\rm sat}}|{\rm -y}\rangle_j\right) \Bigg] |{\rm -x}\rangle_c\\ \nonumber
    &N_{\rm sat}=4n+3 : \vert \psi(t=3.5T) \rangle = \dfrac{1}{\sqrt{2}}\Bigg[\left(\prod_{j=1}^{N_{\rm sat}}|{\rm +y}\rangle_j\right)+i\left(\prod_{j=1}^{N_{\rm sat}}|{\rm -y}\rangle_j\right)\Bigg] |{\rm -x}\rangle_c
\end{align}
Finally at $t=4T$, both for $N_{\rm sat}=4n+1$ and $N_{\rm sat}=4n+3$, the system evolves to the state:
\begin{equation}
    \vert \psi(t=4T) \rangle = \dfrac{1}{\sqrt{2}}\Bigg[\left(\prod_{j=1}^{N_{\rm sat}}|{\rm +z}\rangle_j\right)-\left(\prod_{j=1}^{N_{\rm sat}}|{\rm -z}\rangle_j\right) \Bigg] |{\rm -x}\rangle_c,
\end{equation}
as mentioned in Eq.~5 in the main text. Thus, the central spin magnetization oscillates with a period of $8T$, when $N_{\rm sat}$ is odd. However, the satellite spin magnetization oscillates with a period of $24T$. To see this, we note that at $t=6T$, the system evolves to the Bell-cat state:
\begin{equation}
    \vert \psi(t=6T) \rangle = \frac{1}{\sqrt{2}} \Bigg[\left(\prod_{j=1}^{N_{\rm sat}}\vert {\rm +x} \rangle_j \right)\vert {\rm +x} \rangle_c+i \left(\prod_{j=1}^{N_{\rm sat}}\vert {\rm -x} \rangle_j \right)\vert {\rm -x} \rangle_c \Bigg]
\end{equation}
Analogous to our analysis for the even $N_{\rm sat}$ case, we conclude that the magnetization oscillates with a period of $24T$. These results are corroborated in the magnetization and entanglement oscillation plots given in Fig.~3(b) of the main text.

\begin{figure}[t]
    \centering
    \includegraphics[width=0.92\linewidth]{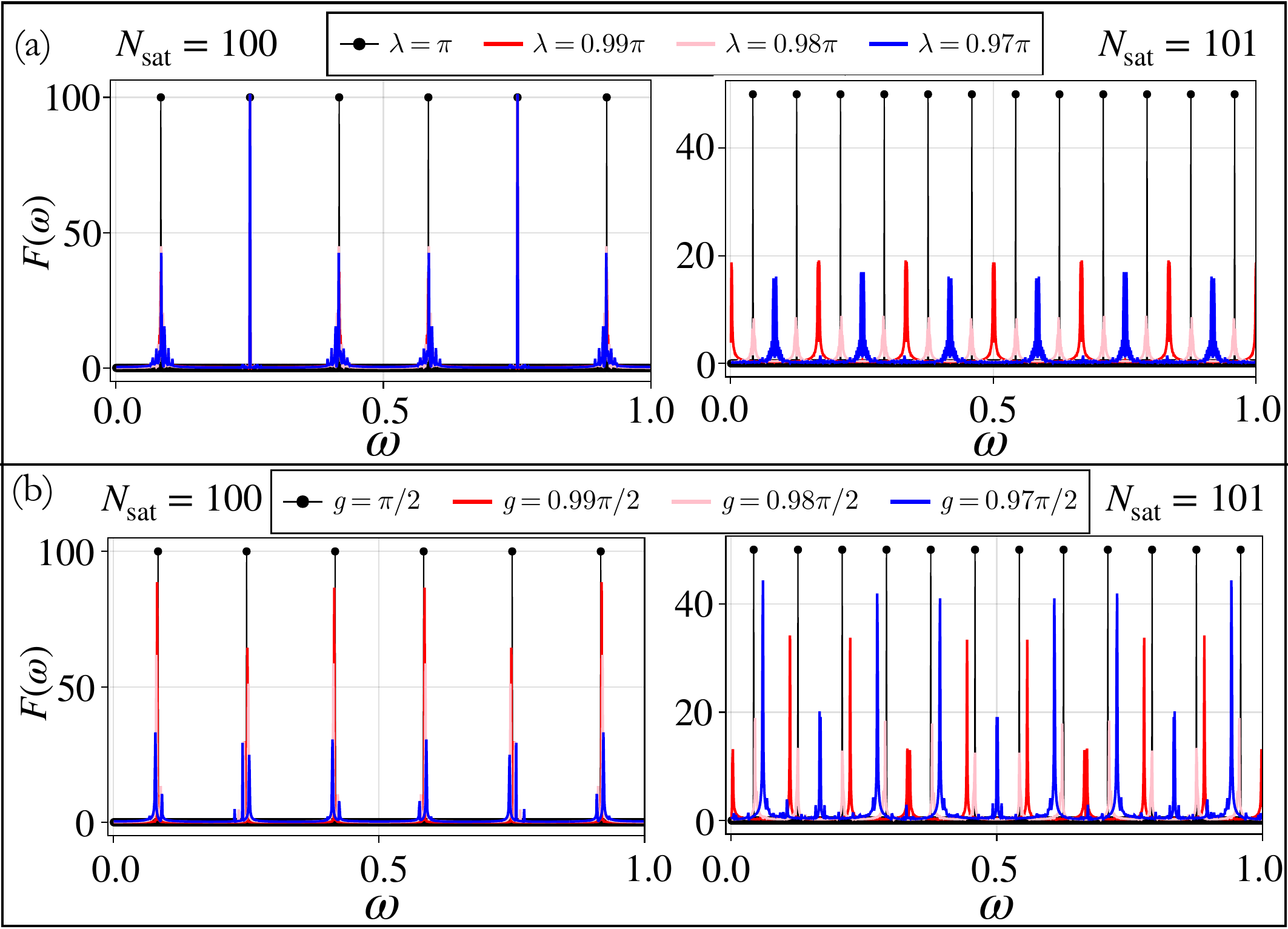}
    \caption{The parity dependence of the higher-period spectral response away from $\lambda=\pi$ and $g=\pi/2$ for $N_{\rm sat}=100$ and $101$ over a finite time window. The Fourier transform of the stroboscopic magnetization is computed over 1200 drive periods for $g=\pi/2$ with varying $\lambda$ (top panel), and $\lambda=\pi$ with varying $g$ (bottom panel). In both cases, the position of the Fourier peaks remains pinned for $N_{\rm sat}=100$, and it shifts and broadens for $N_{\rm sat}=101$. These results demonstrate the parity dependence of the spectral response.}
    \label{fig:response}
\end{figure}

At this point, it is instructive to examine the time-evolution of the generators of the stabilizer group. The stabilizer group generators for the x-polarized initial state are $\mathcal{G}_c = \sigma_c^x$ and $\mathcal{G}_s^{(i)} = \sigma_i^x \, \, \forall i \in \{1,N_{\rm sat}\}$. As discussed above, after $m$ drive periods, the stabilizer generator of the central and satellite spins takes the form, $\mathcal{G}_c (mT) = (-1)^a \sigma_c^{\alpha} \prod_{i=1}^{N_{\rm sat}} \sigma_i^{\beta}$ and $\mathcal{G}_s^{(j)} (mT) = (-1)^b \sigma_c^{\gamma} \sigma_j^{\delta} \prod_{i\ne j} \sigma_i^{\epsilon}$, where $a, b \in \{0,1\}$ and $\alpha, \beta, \gamma, \delta, \epsilon \in \{0,x,y,z\}$, and $\sigma^{0}$ represents the $2 \times 2$ Identity matrix. The time-evolution of the stabilizer generators over 12 drive periods is presented in Tab.~\ref{tab:stabilizer_cycle}. The full stabilizer-evolution can then be obtained using the following rules:
\begin{eqnarray}
\mathcal{G}_c\bigg((m+4)T\bigg) &=& -\mathcal{G}_c(mT), \,\,\mathcal{G}_c\bigg((m+8)T\bigg) = \mathcal{G}_c(mT)\\
\mathcal{G}_s\bigg((m+12)T\bigg) &=& -\mathcal{G}_s(mT), \,\, \mathcal{G}_s\bigg((m+24)T\bigg) = \mathcal{G}_s(mT)
\end{eqnarray}
These results provide a comprehensive characterization of the nature of the entangled states generated during the time-evolution of the system at the Clifford point.

\begin{figure*}[t]
    \includegraphics[width=0.95\textwidth]{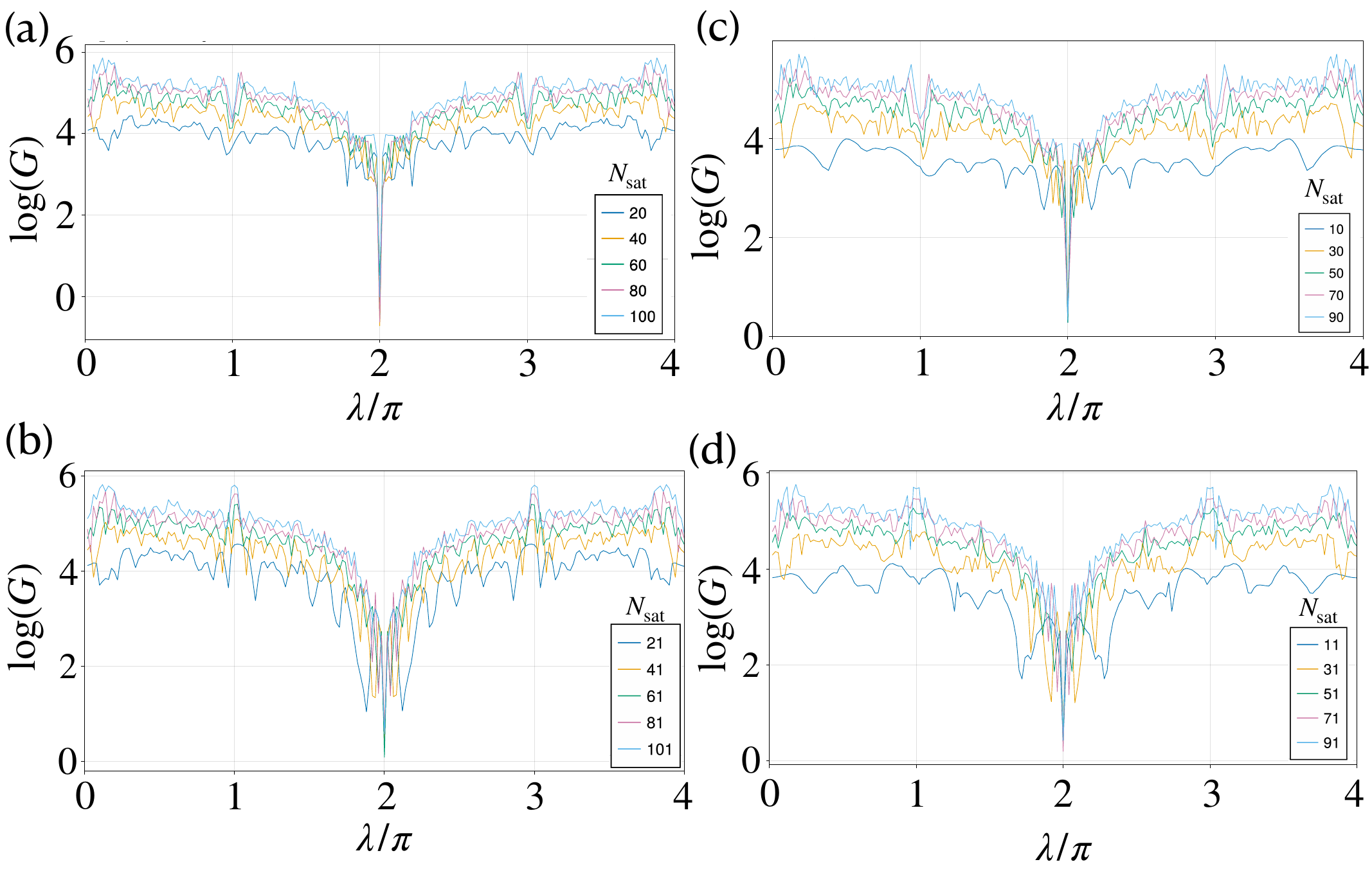}
    \caption{Dependence of the QFI, $G (t=100T)$ (Eq.~6 of the main text) on $\lambda$ for $N_{\rm sat}=$ $4n$ (a), $4n+1$ (b), $4n+2$ (c), and $4n+3$ (d), when $g=\pi/2$. While the chaotic regions and Clifford points are both associated with a high value of $G$, the chaotic regions exhibit oscillatory behavior that makes it difficult to obtain a system-size dependence; the Clifford points, on the other hand, exhibit a clear system-size dependence as mentioned in the main text.}
    \label{fig:QFIsupp}
\end{figure*}

Before concluding this section, it is worth addressing two additional points. First, as noted in the main text, the spectral response of the system away from the Clifford points exhibits a parity dependence. This is illustrated in Fig.~\ref{fig:response}, where the Fourier response remains pinned for even $N_{\rm sat}$ over a finite time window of 1200 drive periods, but shifts and broadens for odd $N_{\rm sat}$. The other point concerns our discussion of multiparameter metrology in the main text. There, we found that both the Clifford points and the chaotic parameter regimes were associated with high values of $G$ (defined in Eq.~6 of the main text). Furthermore, we demonstrated that the entanglement oscillations at the Clifford points exhibit a QFI scaling of $G (t=nT, N_{\rm sat}) \propto n^{\alpha} N_{\rm sat}^{\beta}$, where $\alpha=2$ for any value of $N_{\rm sat}$ and $\beta=2$ and $1$ for odd and even values of $N_{\rm sat}$, respectively. However, as shown in Fig.~\ref{fig:QFIsupp}, such a systematic dependence cannot be obtained for the chaotic regions due to the oscillatory nature of $G$ in this parameter regime.

\bibliographystyle{apsrev4-2}
\bibliography{ref}